\documentclass[journal]{IEEEtran}

\usepackage{ifpdf}
\usepackage{cite}
\usepackage[pdftex]{graphicx}
\usepackage{enumerate}
\usepackage{amsmath}
\usepackage{bm}
\usepackage{array}
\usepackage{multirow}
\usepackage{algorithm}
\usepackage{algorithmic}
\usepackage{float}
\usepackage{url}
\usepackage{subfigure}
\usepackage{amssymb}
\usepackage{verbatim}
\usepackage{caption}
\usepackage{cite}
\usepackage{bm}
\begin{document}

\title{
JPEG Steganography with Embedding Cost Learning and Side-Information Estimation}

\author{Jianhua~Yang,~
         Yi~Liao,~
         Fei Shang,~
         Xiangui~Kang,~
         Yun-Qing~Shi
\thanks{J. Yang, Y. Liao, F. Shang and X. Kang are with Guangdong Key Lab of Information Security, Sun Yat-sen University, Guangzhou 510006, China (e-mail: isskxg@mail.sysu.edu.cn).}

\thanks{Y. Shi is with Department of ECE, New Jersey Institute of Technology, Newark, NJ 07102, USA (e-mail:shi@njit.edu).}

}

\maketitle

\begin{abstract}

A great challenge to steganography has arisen with the wide application of steganalysis methods based on convolutional neural networks (CNNs).
To this end, embedding cost learning frameworks based on generative adversarial networks (GANs) have been proposed and achieved success for spatial steganography.
However, the application of GAN to JPEG steganography is still in the prototype stage;
its anti-detectability and training efficiency should be improved.
In conventional steganography, research has shown that the side-information calculated from the precover can be used to enhance security.
However, it is hard to calculate the side-information without the spatial domain image.
In this work, an embedding cost learning framework for JPEG Steganography via a Generative Adversarial Network (JS-GAN) has been proposed, the learned embedding cost can be further adjusted asymmetrically according to the estimated side-information.
Experimental results have demonstrated that the proposed method can automatically learn a content-adaptive embedding cost function, and use the estimated side-information properly can effectively improve the security performance. For example, under the attack of a classic steganalyzer GFR with quality factor 75 and 0.4 bpnzAC, the proposed JS-GAN can increase the detection error 2.58\% over J-UNIWARD, and the estimated side-information aided version JS-GAN(ESI) can further increase the security performance by 11.25\% over JS-GAN.
\end{abstract}


\section{INTRODUCTION}
JPEG steganography is a technology that aims at covert communication, through modifying the coefficients of the Discrete Cosine Transform (DCT) of an innocuous JPEG image.
By restricting the modification to the complex area or the low-frequency area of the DCT coefficients, content-adaptive steganographic schemes guarantee satisfactory anti-detection performance, which has become a mainstream research direction. Since the Syndrome-trellis codes (STC) \cite{filler2011minimizing} can embed a given payload with a minimal embedding impact, current research on content-adaptive steganography has mainly focused on how to design a reasonable embedding cost function \cite{holub2014universal,UERD}.

In contrast, steganalysis methods try to detect whether the image has secret messages hidden in it. Conventional steganalysis methods are mainly based on statistical features with ensemble classifiers \cite{song2015steganalysis,holub2014low,FLD-ensemble}.
To further improve the detection performance, the selection channel has been incorporated for feature extraction \cite{ denemark2016steganalysis,tang2015adaptive}. In recent years, steganalysis methods based on Convolutional Neural Networks (CNNs) have been researched, initially in the spatial domain \cite{TanStacked2014,QianDeep2015,Yedroudj,YeDeep2017}, and have also achieved success in the JPEG domain \cite{zeng2016large,xu2017deep,yang2018jpeg,boroumand2018deep}.
Current research has shown that a CNN-based steganalyzer can reduce the detection error dramatically compared with conventional steganalyzers, and the security of conventional steganography faces great challenges.

With the development of deep neural networks, recent work has proposed an automatic stegagography method to jointly train the encoder and decoder networks. The encoder can embed the message and generate an indistinguishable stego image. Then the message can be recovered from the decoder. However, these methods mainly use the image as the information due to the visual redundancy, and can not guarantee to recover hidden binary message accurately \cite{HiDDeN,ANIPS}.

In the conventional steganographic scheme, the embedding cost function is designed heuristically, and the measurement of the distortion can not be adjusted automatically according to the strategy of the steganalysis algorithm. Frameworks for automatically learning the embedding cost by adversarial training have been proposed, and can achieve better performance than the conventional method in the spatial domain \cite{tang2017automatic,yang2019embedding,tang2020automatic}. In \cite{tang2017automatic} there was proposed an automatic steganographic distortion learning framework with generative adversarial networks (ASDL-GAN), which can generate an embedding cost function for spatial steganography.
UT-GAN (U-Net and Double-tanh function with GAN based framework)\cite{yang2019embedding} further enhanced the security performance and training efficiency of the GAN-based steganographic method by incorporating a U-Net \cite{ronneberger2015u} based generator and a double-tanh embedding simulator. The influence of high pass filters in the pre-processing layer of the discriminator were also investigated. Experiments show that UT-GAN can achieve a better performance than the conventional method. SPAR-RL (Steganographic Pixel-wise Actions and Rewards with Reinforcement Learning) \cite{tang2020automatic} uses reinforcement learning to improve the loss function of GAN based steganography, and experimental results show that it further improves the stable performance. Although embedding cost learning has been developed in the spatial domain, it is still in its initial stages in the JPEG domain. In our previous conference presentation \cite{yang2019towards}, we proposed a JPEG steganography framework based on GAN to learn the embedding cost. Experimental results show that it can learn the adaptivity and achieve a performance comparable with the conventional methods.

To preserve the statistical model of the cover image, some steganography methods use the knowledge of the so-called precover \cite{ker2007fusion}.
The precover is the original spatial image before JPEG compression.
In conventional methods, previous studies have shown that using asymmetric embedding, namely, with +1 and -1 processes with different embedding costs, can further improve the security performance.
Side-informed JPEG steganography calculates the rounding error of the DCT coefficients with respect to the compression step of the precover, then uses the rounding error as the side-information to adjust the embedding cost for asymmetric embedding \cite{denemark2015side}.
However, it is hard to obtain the side-information without the precover, so the researcher tries to estimate the precover to calculate the estimated side-information.
In \cite{li2019jpeg}, a precover estimation method that uses series filters was proposed.
The experimental results show that although it is hard to estimate the amplitude of the rounding error, the security performance can be improved by using the polarity of the estimated rounding error.

Although initial studies of the automatic embedding cost learning framework have shown that it can be content-adaptive, its security performance and training efficiency should be improved.
In conventional steganography, the side-information estimation is heuristically designed, and
the precision of the estimation depends on experience and experiments.
How to estimate the side-information through CNN and adjust the embedding cost asymmetrically with the unprecise side-information needs to be investigated.

In the present paper, we extend \cite{yang2019towards}, and the learned embedding cost can be further adjusted asymmetrically according to the estimated side-information. The main contributions of this paper can be summarized as follows.

\begin{enumerate}[1)]
	\item  We further develop the GAN-based method of generating an embedding cost function for JPEG
          steganography. Unlike conventional hand-crafted cost functions, the proposed method can automatically learn the embedding cost via adversarial training.

   \item To solve the gradient-vanishing problem of the embedding simulator, we propose a  gradient-descent friendly embedding simulator to generate the modification map with a higher efficiency.

  \item Under the condition of lacking the uncompressed image, we propose a CNN-based side-information estimation method. The estimated rounding error has been used as the side-information for asymmetrical embedding to further improve the security performance.

\end{enumerate}

The rest of this paper is organized as follows.
In Section II, we briefly introduce the basics of the proposed steganographic algorithm, which includes the concept of distortion minimization framework and side-informed JPEG steganography.
A detailed description of the proposed GAN-based framework and  side-information estimation method is given in Section III.
Section IV presents the experimental setup and verifies the adaptivity of the proposed embedding scheme.
The security performance of our proposed steganography under different payloads compared with the conventional methods is also shown in Section IV.
Our conclusions and avenues for future research are presented in Section V.

\section{Preliminaries}

\subsection{Notation}
In this article, the capital symbols stand for matrices, and $i,j$ are used to index the elements of the matrices.
The symbols ${\rm\mathbf{C}}=(C_{i,j}),{\rm\mathbf{S}}=(S_{i,j}) \in \mathbb{R}^{h \times w}$  represent the 8-bit grayscale cover image and its stego image of size $h \times w$ respectively, where $S_{i,j} =\{\max(C_{i,j}-1,0), C_{i,j}, \min(C_{i,j}+1, 255)\}$.
${\rm\mathbf{P}}=(p_{i,j})$ denotes the embedding probability map.
${\rm\mathbf{N}}=(n_{i,j})$ stands for a random matrix with uniform distribution ranging from 0 to 1.
${\rm\mathbf{M}}=(m_{i,j})$ stands for the modification map, where $m_{i,j} \in \{-1, 0, 1\}$.

\subsection{Distortion minimization framework}
Most of the successful steganographic methods embed the payload by obeying a distortion minimization rule that the sender embeds a payload of $m$ bits while minimizing the average distortion \cite{filler2011minimizing},
\begin{equation}
	\arg \min_{\pi} E_\pi[D] = \sum_{\bm{S}} \pi(\bm{S})D(\mathbf{S}),
	\label{minimizing_dis}
\end{equation}
\begin{equation}
	\text{subject to} \  m= - \sum_{\mathbf{S}} \pi(\mathbf{S})log(\pi(\mathbf{S})),
	\label{payload_constrain}
\end{equation}
where $\pi(\bm{S})$ stands for the modification distribution of modifying ${\rm\mathbf{C}}$ to ${\rm\mathbf{S}}$, $D(\mathbf{S})$ is the distortion function that measures the impact of embedding modifications and is defined as:
\begin{equation}
	D(\mathbf{S}) \triangleq  D(\mathbf{C},\mathbf{S}) = \sum_{i=1}^{h}\sum_{j=1}^{w} \rho_{i,j} \left| S_{i,j} - C_{i,j} \right|,
\end{equation}
where $\rho_{i,j}$ represents the cost of replacing the $C_{i,j}$ with $S_{i,j}$.
In most symmetric embedding schemes, the costs of increasing or decreasing $C_{i,j}$ by 1 are equal during embedding, \textit{i.e.}, $\rho^{+1}_{i,j} = \rho^{-1}_{i,j} = \rho_{i,j}$.
The optimal solution of Eq.~\ref{minimizing_dis} has the form of a Gibbs distribution \cite{filler2010gibbs},
\begin{equation}
	\pi(\bm{S}) = \frac{ exp(-\lambda D(\bm{S}))}{\sum_{\bm{S}} exp(-\lambda D(\bm{S}))},
\end{equation}
the scalar parameter $\lambda > 0$ needs to be calculated from the payload constraint in Eq.~\ref{payload_constrain}.

\subsection{Side-informed JPEG steganography}

Side-informed (SI) JPEG steganography uses the additional message to adjust the cost for better embedding. In \cite{denemark2015side}, the rounding error of the DCT coefficient is calculated from the precover, then the embedding cost is adjusted by using the rounding error as the side-information for an asymmetric embedding.
For a given precover, the rounding error is defined as follows:
\begin{equation}
   e_{ij}= U_{i,j}-C_{i,j},
\end {equation}
where $U_{i,j}$ is the non-rounded DCT coefficient, and $C_{i,j}$ is the rounded DCT coefficient of the cover image.
When generating the stego $\rm{\mathbf{S}}$ by using the ternary embedding steganography with side-information, the embedding cost $\rho _{ij}$ is calculated at first, then the costs of changing $C_{ij}$ by $ \pm {\rm sign}(e_{ij})$ can be adjusted as follows:
\begin{equation}
\begin{cases}
	\rho _{ij}^{(SI)+}= (1-2|e_{ij}|) \rho _{ij}& {\rm if\ } S_{i,j}=C_{ij}+{\rm sign}(e_{ij})  \\
	\rho _{ij}^{(SI)-}=\rho _{ij}& {\rm if\ } S_{i,j}=C_{ij}-{\rm sign}(e_{ij})
\end{cases}.
\label{cost_adjust}
\end{equation}

When the precover is unavailable, \cite{li2019jpeg} tried to estimate the side-information from the precover by first using series filters, then calculating the estimated rounding error $\hat{e}$ to be the side-information with which to adjust the embedding cost.

\begin{equation}
\begin{cases}
	\rho _{i,j}^{(ESI)+}=g(\hat{e}_{i,j})\cdot \rho _{i,j}& {\rm if\ } S_{i,j}=C_{i,j}+{\rm sign}(\hat{e}_{i,j})  \\
	\rho _{i,j}^{(ESI)-}=\rho _{i,j}& {\rm if\ } S_{i,j}=C_{i,j}-{\rm sign}(\hat{e}_{i,j})
\end{cases},
\label{cost_adjust}
\end{equation}

\begin{equation}
g(\hat{e}_{i,j})=
\begin{cases}
	1-2\left | \hat{e}_{i,j} \right | & {\rm if\ } \left | \hat{e}_{i,j} \right |\leq 0.5 \\
	\eta & otherwise
\end{cases},
\label{eq_esi_adjust}
\end{equation}
where $\eta$ is used to make sure the embedding cost is positive when the absolute value of
the side-information is greater than 0.5.
It should be noted that steganography with estimated side-information is even inferior to the methods without side-information due to the imprecision in the amplitude of the rounding error.
To solve this problem, the authors proposed a method using polarity to adjust the embedding cost. The sign of the side-information is used to adjust the cost and the amplitude is ignored.

\begin{equation}
\begin{cases}
	\rho _{i,j}^{(ESI)+}= \eta .\rho _{i,j}& {\rm if\ } S_{i,j}=C_{i,j}+{\rm sign}(\hat{e}_{i,j})  \\
	\rho _{i,j}^{(ESI)-}=\rho _{i,j}& {\rm if\ } S_{i,j}=C_{i,j}-{\rm sign}(\hat{e}_{i,j})
\end{cases}.
\label{cost_adjust}
\end{equation}

\section{The Proposed Cost Function Learning Framework for JPEG Steganography}
In this section, we propose an embedding cost learning framework for JPEG Steganography based on GAN (JS-GAN).
We conduct an estimation of the side-information (ESI) based on CNN to asymmetrically adjust the embedding cost to further improve the security, and the version which uses ESI as a help is referred to as JS-GAN(ESI).

\begin{figure*}
	\centering
	\includegraphics[width=14cm]{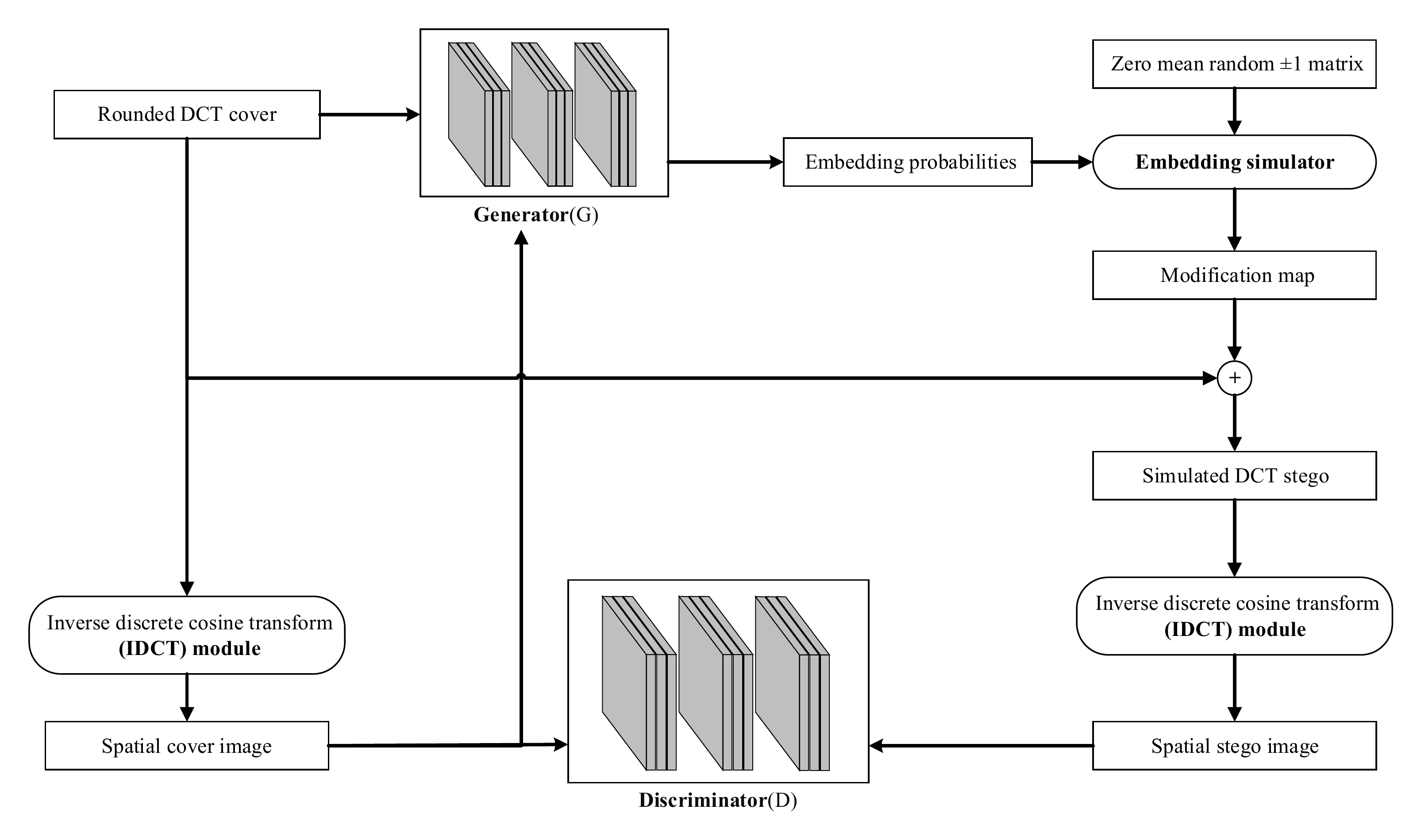}
	\caption{ Architecture of the proposed JS-GAN.}
	\label{fig_framework}
\end{figure*}

\subsection{JS-GAN}
The overall architecture of the proposed JS-GAN is shown in Fig.~\ref{fig_framework}.
It's mainly composed of four modules: a generator, an embedding simulator, an IDCT module, and a discriminator.
The training steps are described by \textbf{Algorithm 1}.

For an input of a rounded DCT matrix $\rm{\mathbf{C}}$, the $p_{i,j} \in [0,1]$ denotes the corresponding embedding probability produced by the adversarially trained generator.
Since the probabilities of increasing or decreasing $C_{i,j}$ are equal, we set $p^{+1}_{i,j} = p^{-1}_{i,j} = p_{i,j}/2$ and the probability that $C_{i,j}$ remains unchanged is $p^{0}_{i,j} = 1 - p_{i,j}$.
We also feed the spatial cover image converted from the rounded DCT matrix into the generator to improve the performance of our method.

The embedding simulator is used to generate the corresponding modification map.
The DCT matrix of the stego image is obtained by adding the modification map to the DCT matrix of the cover image.
By applying the IDCT module \cite{yang2019towards}, we can finally produce the spatial cover-stego image pair.
The discriminator tries to distinguish the spatial stego images from the innocent cover images.
Its classification error is regarded as the loss function to train the discriminator and generator using the gradient descent optimization algorithm.

\begin{algorithm}
	\caption{Training steps of JS-GAN}
	\label{alg:alg1}
	\begin{algorithmic}
		\REQUIRE ~~\\
		
		Rounded DCT matrix of cover image.\\
		Zero mean random $\pm 1$ matrix.\\
		
		\begin{enumerate}[\textbf{Step} 1:]
			\item Input the rounded DCT matrix of the cover image and the corresponding spatial cover image into the generator to obtain the embedding probability.
			\item  Generate modification map using the proposed embedding simulator.
			\item  Add the modification map into the DCT coefficients matrix of the cover image to generate the DCT coefficients matrix of the stego image.
			\item Convert the cover and stego DCT coefficients matrix to spatial image by using the IDCT module.
			\item Feed the spatial cover-stego pair into the discriminator to obtain the loss of generator and discriminator.
			\item Update the parameters of the generator and discriminator alternately, using the gradient descent optimization algorithm of Adam \cite{kingma2014adam} to minimize the loss.
		\end{enumerate}
		
	\end{algorithmic}
\end{algorithm}

After training the JS-GAN as shown in \textbf{Algorithm 1} with 0.5 bpnzAC (bit per non-zero AC DCT coefficient) payload for a certain number of iterations, the trained generator is capable of generating an embedding probability, which is then used for the follow-up steganography.
Since the embedding cost should be constrained to $ 0 \le \rho_{i,j}\le \infty$,  it can be computed based on the embedding probability $p_{i,j}$ as follows \cite{sedighi2016content}:
\begin{equation}
	\rho_{i,j} = \ln(2/p_{i,j}-1).
	\label{eq_prob2cost}
\end{equation}

To further improve the performance, the embedding costs from the same location of DCT blocks have been smoothed by a Gaussian filter as a post process\cite{li2014investigation}. The incorporation of this Gaussian filter will improve the security performance by about 1.5\%.
After designing the embedding cost, the STC encoder \cite{filler2011minimizing} has been applied to embed the specific secret messages to generate the actual stego image.
Detailed descriptions of each module of JS-GAN will be given in the following sections.

\subsubsection{Architecture of the generator G}

\begin {table*}[ht]
\renewcommand\arraystretch{1.0}
\centering
\caption {Configuration details of the generator}
\begin{tabular}{ |p{1.8cm}<{\centering}|p{9cm}<{\centering}|p{2cm}<{\centering}|p{2.5cm}<{\centering}| }
	\hline
	Group/Layer & Process& Kernel size  & Output size              \\
	\hline
	Input & Concatenation of the input DCT and corresponding spatial image      & / & $2\times(256 \times 256)$  \\
	\hline
	Group 1 &  Convolution-Batch Normalization-Leaky ReLU  & $16\times(3\times3)$      & $16\times(128 \times 128)$ \\
	\hline
	Group 2 & Convolution-Batch Normalization-Leaky ReLU & $32\times(3\times3)$ & $32\times(64 \times 64)$ \\
	\hline
	Group 3 & Convolution-Batch Normalization-Leaky ReLU & $64\times(3\times3)$& $64\times(32 \times 32)$  \\
	\hline
	Group 4 &Convolution-Batch Normalization-Leaky ReLU&  $128\times(3\times3)$ &$ 128\times(16 \times 16) $ \\
	\hline
	Group 5 & Convolution-Batch Normalization-Leaky ReLU& $128\times(3\times3)$  &$ 128\times(8 \times 8)$  \\
	\hline
	Group 6 & Convolution-Batch Normalization-Leaky ReLU& $128\times(3\times3)$ & $128\times(4 \times 4) $   \\
	\hline
	Group 7& Convolution-Batch Normalization-Leaky ReLU &   $128\times(3\times3)$& $128\times(2 \times 2)$  \\
	\hline
	Group 8 & Convolution-Batch Normalization-Leaky ReLU & $128\times(3\times3)$ & $128\times(1 \times 1) $ \\
	\hline
	Group 9 & Deconvolution-Batch Normalization-ReLU &  $128\times(5\times5)$ & $128\times(2 \times 2)$ \\
	\hline
	C1      & Concatenation of the feature maps from Group 7 and Group  9 &/& $256\times(2 \times 2)$  \\
	\hline
	Group 10 &Deconvolution-Batch Normalization-ReLU&   $128\times(5\times5)$ & $128\times(4 \times 4)$ \\
	\hline
	C2 &  Concatenation of the feature maps from Group 6 and Group  10     &/ & $256\times(4 \times 4)$\\
	\hline
	Group 11&Deconvolution-Batch Normalization-ReLU &  $128\times(5\times5)$& $128\times(8 \times 8) $  \\
	\hline
	C3 &Concatenation of the feature maps from Group 5 and Group  11&  /      & $256\times(8 \times 8)$\\
	\hline
	Group 12 & Deconvolution-Batch Normalization-ReLU &  $128\times(5\times5)$ & $128\times(16 \times 16)$  \\
	\hline
	C4 & Concatenation of the feature maps from Group 4 and Group  12&  /      & $256\times(16 \times 16)$ \\
	\hline
	Group 13 & Deconvolution-Batch Normalization-ReLU&   $64\times(5\times5)$& $64\times(32 \times 32)$ \\
	\hline
	C5 & Concatenation of the feature maps from Group 3 and Group  13&  /      & $128\times(32 \times 32)$ \\
	\hline
	Group 14 & Deconvolution-Batch Normalization-ReLU& $32\times(5\times5)$ & $32\times(64 \times 64) $ \\
	\hline
	C6 & Concatenation of the feature maps from Group 2 and Group  14&  /      & $64\times(64 \times 64)$ \\
	\hline
	Group 15 & Deconvolution-Batch Normalization-ReLU& $16\times(5\times5)$ & $16\times(128 \times 128) $ \\
	\hline
	C7 & Concatenation of the feature maps from Group 1 and Group  15&  /& $32\times(128 \times 128)$  \\
	\hline
	Group 16& Deconvolution-Batch Normalization & $1\times(5\times5)$ & $1\times(256 \times 256)$  \\
	\hline
	Output & Sigmoid  &/ & $1\times(256 \times 256)$  \\
	\hline	
\end{tabular}
\label{table_gen}
\end{table*}

The main purpose of JS-GAN is to train a generator that can generate the embedding probability with the same size as the DCT matrix of the cover image, and the process step can be regard as an image-to-image translation task.
Based on the superior performance in image-to-image translation and high training efficiency, we use U-Net \cite{ronneberger2015u} as a reference structure to design our generator.
The generator of JS-GAN contains a contracting path and an expansive path with 16 groups of layers.
The contracting path is composed of 8 operating groups: each group includes a convolutional layer with stride 2 for down-sampling, a batch-normalization layer, and a leaky rectified linear unit (Leaky-ReLU) activation function.
The expansive path consists of the repeated application of the deconvolution layer, each time followed by a batch-normalization layer and ReLU activation function.
After up-sampling the feature map to the same size as the input, the final sigmoid activation function is added to restrict the output within the range of 0 to 1 in order to meet the requirements of being an embedding probability.
To achieve pixel-level learning and facilitate the back-propagation, concatenations of feature maps are placed between each pair of mirrored convolution and deconvolution layers.
The specific configuration of the generator is given in Table \ref{table_gen}.

\subsubsection{The embedding simulator}
As shown in Fig. 1, an embedding simulator is required to generate the corresponding modification map according to the embedding probability.
Conventional steganography methods \cite{holub2014universal,pevny2010using,holub2012designing,li2014new,sedighi2016content}  use a staircase function \cite{fridrich2007practical} to simulate the embedding process:
\begin{equation}
m_{i,j}=
\begin{cases}
	-1, & {\rm if\ } \ n_{i,j}<p_{i,j}/2 \\
	1, & {\rm if\ } \ n_{i,j}>1-p_{i,j}/2\\
	0, & {\rm otherwise\ }
\end{cases},
\label{eq_stair}
\end{equation}
where $m_{i,j}$ is the modification map with values $\pm 1$ and 0, $n_{i,j}$ stands for a random number from the uniform distribution on the interval $[0,1]$, and $p_{i,j}$ is the embedding probability.

Although Eq.~\ref{eq_stair} has been widely used in conventional methods, the staircase function cannot be put into the pipeline of the training phase of GAN because most of the derivatives are zero, which will lead to the gradient-vanishing problem.
In the present paper, we propose an embedding simulator that uses the learned embedding probability.
Since the primary target of the embedding simulator in JS-GAN is to make more modifications of the elements with a higher embedding probability, we use the probability as a modification map. The stego DCT matrix is generated by adding the cover DCT matrix and the corresponding modification map.
It should also be noticed that the learned probabilities range from 0 to 1, which can not simulate a modification with a negative sign.
To solve this problem, our proposed embedding simulator multiplies the embedding probability map with a zero mean random $\pm 1$ matrix to obtain the modification map $m_{i,j}$.
The specific implementation of our proposed embedding simulator is given in Eq.~\ref{eq_embed},
\begin{equation}
m_{i,j}=p_{i,j} \times (1-2 \left[n_{i,j} > 0.5 \right]),
\label{eq_embed}
\end{equation}
where $\left[P\right]$ is the Iverson bracket, \textit{i.e.}, equal to 1 when the statement $P$ is true and 0 otherwise.
The embedding simulator we proposed can generate the corresponding DCT matrix of stego efficiently and is gradient-descent friendly. Experimental results also prove that applying this embedding simulator in our JS-GAN means it can learn an adaptive embedding probability.

\subsubsection{Architecture of the discriminator D}

\begin{table*}
\centering
\caption{Configuration details of the discriminator}
\begin{tabular}{|c|c|l|c|}
\hline
Groups                     & \multicolumn{2}{c|}{Process}                                                                                                     & Output size        \\ \hline
Preprocess layer           & \multicolumn{2}{c|}{Gabor filtering}                                                                                             & $16\times(256\times256)$                  \\ \hline
L1                         & \multicolumn{2}{c|}{Conv($stride$ 1, $k$ = 12)-BN-TLU}                                                                               & $12\times(256\times256)$                  \\ \hline
\multirow{3}{*}{ResBlock1} & \begin{tabular}[c]{@{}c@{}}Conv($stride$ 1, $k$ = 12)-BN-ReLU\\ Conv($stride$ 1, $k$ = 12)-BN\end{tabular}   & Add-ReLU                  & \multirow{3}{*}{$24\times(128\times128)$} \\ \cline{2-3}
& \begin{tabular}[c]{@{}c@{}}Conv($stride$ 1, $k$ = 12)-BN-ReLU\\ Conv($stride$ 2, $k$ = 24)-BN\end{tabular}   & \multirow{2}{*}{Add-ReLU} &                    \\ \cline{2-2}
& Conv($stride$ 2, $k$ = 24)-BN                                                                            &                           &                    \\ \hline
\multirow{3}{*}{ResBlock2} & \begin{tabular}[c]{@{}c@{}}Conv($stride$ 1, $k$ = 24)-BN-ReLU\\ Conv($stride$ 1, $k$ = 24)-BN\end{tabular}   & Add-ReLU                  & \multirow{3}{*}{$48\times(64\times64)$} \\ \cline{2-3}
& \begin{tabular}[c]{@{}c@{}}Conv($stride$ 1, $k$ = 24)-BN-ReLU\\ Conv($stride$ 2, $k$ = 48)-BN\end{tabular}   & \multirow{2}{*}{Add-ReLU} &                    \\ \cline{2-2}
& Conv($stride$ 2, $k$ = 48)-BN                                                                            &                           &                    \\ \hline
\multirow{3}{*}{ResBlock3} & \begin{tabular}[c]{@{}c@{}}Conv($stride$ 1, $k$ = 48)-BN-ReLU\\ Conv($stride$ 1, $k$ = 48)-BN\end{tabular}   & Add-ReLU                  & \multirow{3}{*}{$96\times(32\times32)$} \\ \cline{2-3}
& \begin{tabular}[c]{@{}c@{}}Conv($stride$ 1, $k$ = 48)-BN-ReLU\\ Conv($stride$ 2, $k$ = 96)-BN\end{tabular}   & \multirow{2}{*}{Add-ReLU} &                    \\ \cline{2-2}
& Conv($stride$ 2, $k$ = 96)-BN                                                                            &                           &                    \\ \hline
\multirow{3}{*}{ResBlock4} & \begin{tabular}[c]{@{}c@{}}Conv($stride$ 1, $k$ = 96)-BN-ReLU\\ Conv($stride$ 1, $k$ = 96)-BN\end{tabular}   & Add-ReLU                  & \multirow{3}{*}{$192\times(16\times16)$} \\ \cline{2-3}
& \begin{tabular}[c]{@{}c@{}}Conv($stride$ 1, $k$ = 96)-BN-ReLU\\ Conv($stride$ 2, $k$ = 192)-BN\end{tabular}  & \multirow{2}{*}{Add-ReLU} &                    \\ \cline{2-2}
& Conv($stride$ 2, $k$ = 192)-BN                                                                           &                           &                    \\ \hline
\multirow{3}{*}{ResBlock5} & \begin{tabular}[c]{@{}c@{}}Conv($stride$ 1, $k$ = 192)-BN-ReLU\\ Conv($stride$ 1, $k$ = 192)-BN\end{tabular} & Add-ReLU                  & \multirow{3}{*}{$384\times(8\times8)$} \\ \cline{2-3}
& \begin{tabular}[c]{@{}c@{}}Conv($stride$ 1, $k$ = 192)-BN-ReLU\\ Conv($stride$ 2, $k$ = 384)-BN\end{tabular} & \multirow{2}{*}{Add-ReLU} &                    \\ \cline{2-2}
& Conv($stride$ 2, $k$ = 384)-BN                                                                           &                           &                    \\ \hline
L2                         & \multicolumn{2}{c|}{\begin{tabular}[c]{@{}c@{}}Conv($stride$ 1, $k$ = 384)-BN-ReLU\\ Conv($stride$ 1, $k$ = 384)-BN-ReLU\end{tabular}}   & $384\times(8\times8)$                  \\ \hline
Output layer               & \multicolumn{2}{c|}{Fully Connected-Softmax}                                                                     & 2                  \\ \hline
\end{tabular}
\label{table_dis}
\end{table*}

In \cite{yang2018jpeg}, we used a densenet based steganalyzer as the discriminator.
Because the feature concatenation consumes more GPU memory and training time than the add process, we used a refined residual networks based J-XuNet \cite{xu2017deep} as the discriminator.
Details of the discriminator are shown in Table \ref{table_dis}.
To against Gabor filter based discriminator, a pre-processing layer that incorporates 16 Gabor high pass filters is used to implement a convolution operation to the input image.
Gabor high pass filters can effectively enhance the steganographic signal and help subsequent neural networks extract steganographic features better.
The Gabor filters are defined as follows:
\begin{equation}
G_{\lambda ,\theta ,\phi ,\sigma ,\gamma }(x,y)=e^{\frac{u^{2}+\gamma ^{2} v^{2}}{2\sigma ^{2}}} \cos (2\pi\frac{u }{\lambda }+\phi ),
\end{equation}
where $u=xcos\theta  + ysin\theta$, $v=xsin\theta  + ycos\theta$. $\sigma$ is the standard deviation of gaussian factor and $\lambda$ is the wavelength, $\sigma = 0.56 \lambda$ meanwhile $\sigma=\left \{ 0.75,1 \right \}$.
Directional coefficient $\theta =\left \{ 0,\frac{\pi}{4},\frac{2\pi}{4},\frac{3\pi}{4} \right \}$, phase offset $\phi =\left \{ 0,\frac{\pi}{2} \right \}$, and spatial aspect ratio $\gamma$ = 0.5.

After the preprocessing layer, a convolutional block that consists of a Truncated Linear Unit (TLU) \cite{ye2017deep} is used. The TLU limits the numerical range of the feature map to $(-T, T)$ to prevent large values of input noise from influencing unduly the weight of the deep network, and we set $T=8$. After that, five residual blocks are used to extract features. The structure of each residual block is a sequence consisting of convolution layers, batch-normalization layers, and ReLU activation functions. Finally, after two convolutional layers and a fully connected layer, the networks produce a classification probability from a softmax layer.

\subsubsection{The loss function}
In JS-GAN, the loss function has two parts: the loss of the discriminator and that of the generator.
The goal of the discriminator is to distinguish between the cover image and the corresponding stego image, thus the cross-entropy loss of the discriminator is defined as:

\begin{equation}
l_{D} = - \sum_{i=1}^{2}z_i'\log(z_i),
\label{eq_lossD}
\end{equation}
where $z_1$ and $z_2$ are the softmax outputs of the discriminator, while $z_1'$ and $z_2'$ stand for the ground truth labels.

In addition to the adversarial training of the generator and discriminator, the embedding capacity that determines the payload of the stego image should also be brought into consideration, which can ensure that enough information is embedded.
The loss function of the generator contains two parts: the adversarial loss $l_G^1$ aims to improve the anti-detectability while the entropy loss $l_G^2$ guarantees the embedding capacity of the stego image, and we define them in Eq.~\ref{eq_lossg1} and Eq.~\ref{eq_lossg2} respectively.

\begin{equation}
l_G^1 = -l_D,
\label{eq_lossg1}
\end{equation}

\begin{equation}
l_G^2 =(C_a-\epsilon\times q)^2,
\label{eq_lossg2}
\end{equation}
\begin{equation}
\begin{split}
C_a = \sum_{i=1}^{h}\sum_{j=1}^{w} [-p^{-1}_{i,j}\log_2p^{-1}_{i,j} -p^{+1}_{i,j}\log_2p^{+1}_{i,j}\\-(1-p^{0}_{i,j})\log_2(1-p^{0}_{i,j})],
\end{split}
\end{equation}
where $C_a$ is the embedding capacity, $h$ and $w$ are the height and width of the image respectively, $\epsilon$ denotes the number of non-zero AC coefficients, and $q$ is the target payload.
The total loss for the generator G is the weighted average of these two kinds of losses:

\begin{equation}
l_G = \alpha \times l_G^1 + \beta \times l_G^2,
\label{eq_lossG}
\end{equation}
where $\alpha=1$ and $\beta= 10^{-7}$.
The setting of $\alpha$ and $\beta$ is based on the magnitudes of $l_G^1$ and $l_G^2$.

\subsection{Side-Information estimation aided JPEG steganography}
In this part, we will introduce the CNN-based side-information architecture and the strategy to asymmetrically adjust the embedding cost according to the estimated side-information.
\subsubsection{Side-Information estimation}

\begin{figure*}
\centering
\subfigure[Training the CNN based precover estimation model.]{
	\begin{minipage}[t]{1\linewidth}
		\centering
		\includegraphics[width=12cm]{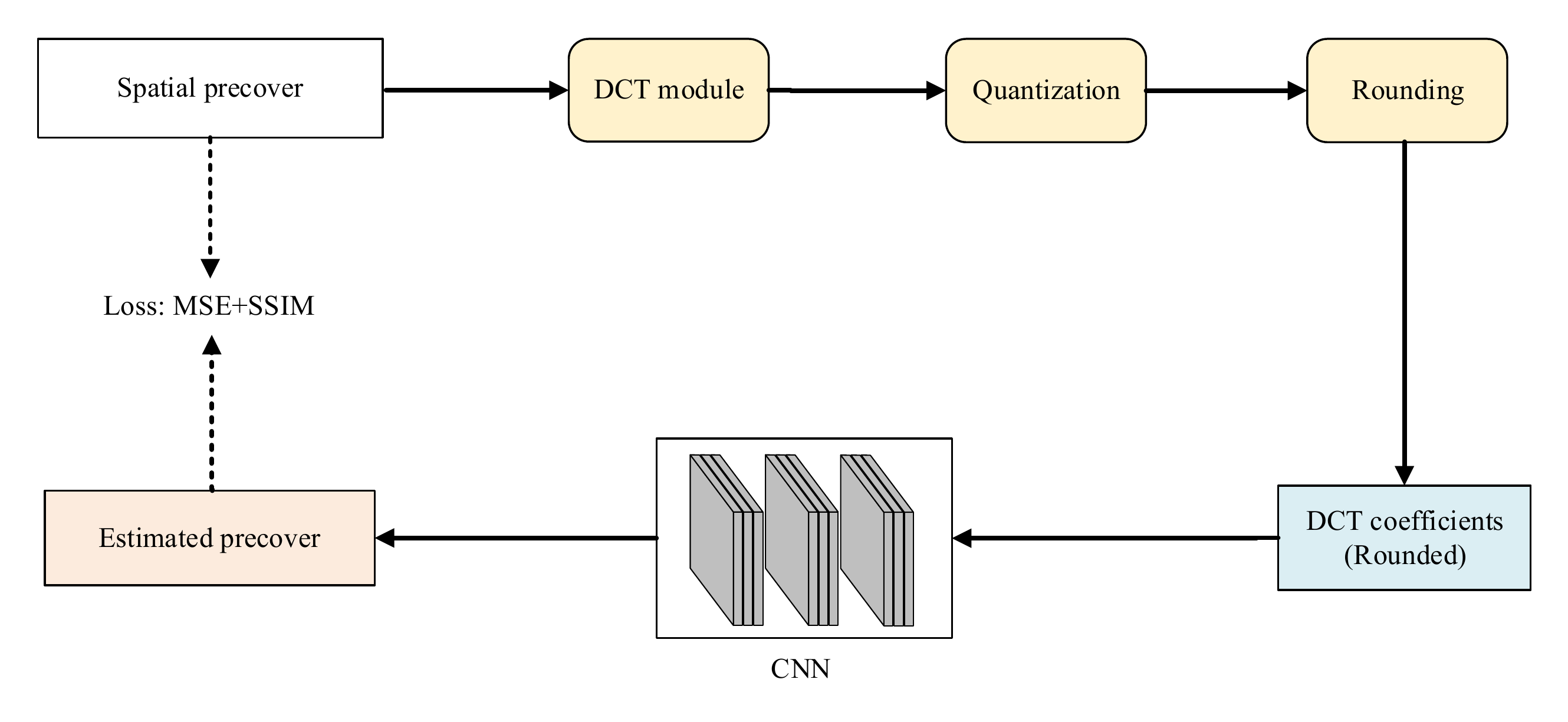}
	\end{minipage}
}
\subfigure[Calculating the estimated side-information.]{
	\begin{minipage}[t]{1\linewidth}
		\centering
		\includegraphics[width=12cm]{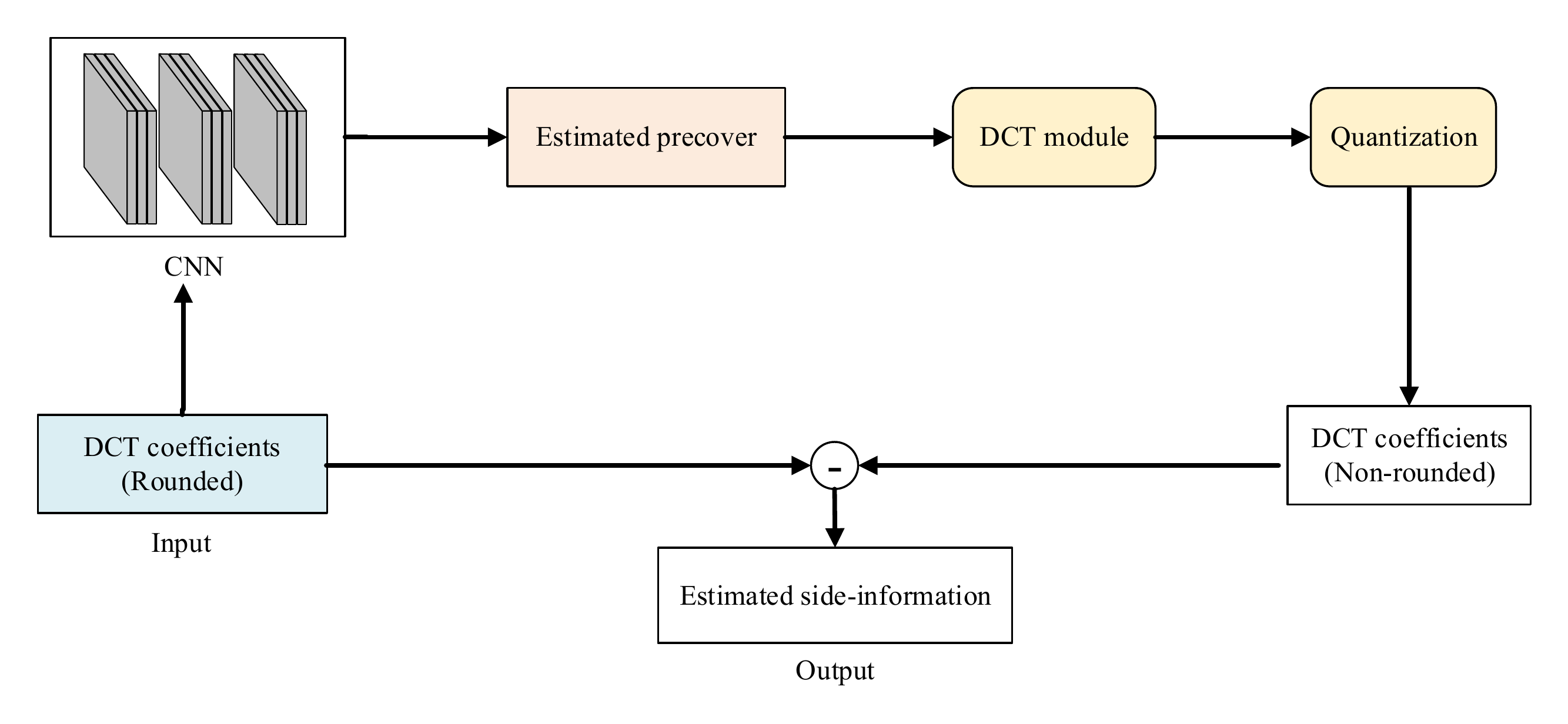}
	\end{minipage}
}
\caption{The architecture of side-information estimation.}
\label{fig_esi_flowchart}
\end{figure*}

The architecture of the side information estimation is shown in Fig. \ref{fig_esi_flowchart}. It is composed of two steps: training the CNN based precover estimation model, and calculating the side-information using the trained model.

The training steps are shown in Fig. \ref{fig_esi_flowchart}(a). For a given spatial precover, the quantized and rounded DCT coefficients are calculated at first, then the DCT coefficients are input to the CNN to obtain the estimated precover.
The loss function of mean square error (MSE) and structural similarity index measure (SSIM) is used to reduce the difference between the estimated precover and the existed spatial precover.
\begin{figure*}
\centering
\includegraphics[width=14cm]{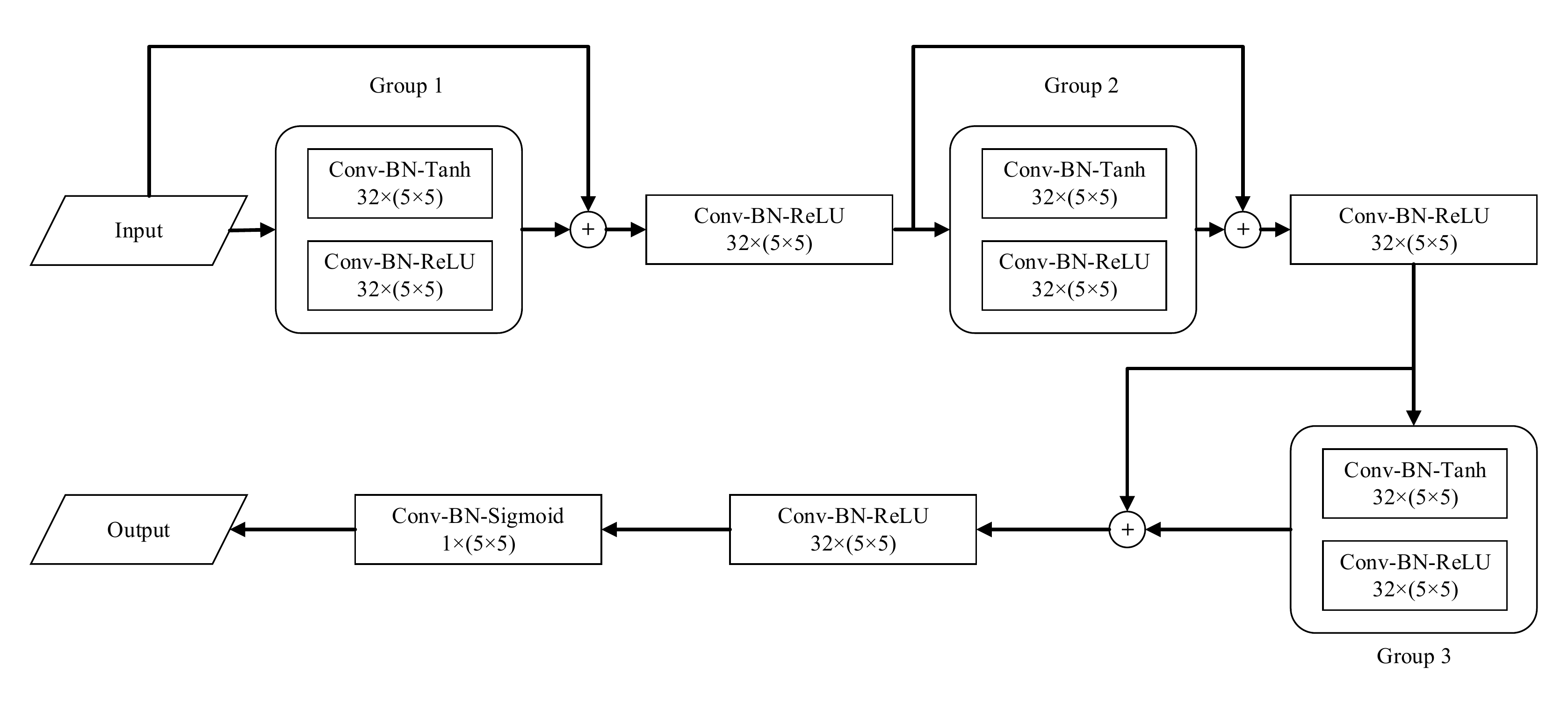}
\caption{The proposed CNN-based precover estimation architecture. Sizes of convolution kernels follow (number of kernels) $\times$ (height $\times$ width).}
\label{fig_esi_network}
\end{figure*}
\begin{equation}
l_{E} = \rm {SSIM} + \rm {MSE},
\label{eq_lossEsi}
\end{equation}
\begin{equation}
{\rm {MSE}}(x,y)=\frac{1}{k}\sum_{i=1}^{k}(x_{i}-y_{i})^{2},
\end{equation}
\begin{equation}
{\rm {SSIM}}(x,y)=[L(x,y)]^{l}[C(x,y)]^{m}[S(x,y)]^{n},
\end{equation}
where $k$ is the batch size of the precover estimation network. $L(x,y)$, $C(x,y)$, $S(x,y)$ represent brightness comparison, contrast comparison and structural comparison respectively. The parameters $l$ \textgreater\ 0, $m$ \textgreater\ 0, $n$ \textgreater\ 0  are used to adjust the importance of the three respective components. $L(x,y)$, $C(x,y)$, $S(x,y)$ are calculated by
\begin{equation}
L(x,y)=\frac{2\mu _{x}\mu _{y}+C_{1}}{\mu _{x}^{2}+\mu _{y}^{2}+C_{1}},
\end{equation}
\begin{equation}
C(x,y)=\frac{2\theta _{x}\theta _{y}+C_{2}}{\theta _{x}^{2}+\theta _{y}^{2}+C_{2}},
\end{equation}
\begin{equation}
S(x,y)=\frac{\theta _{xy}+C_{3}}{\theta _{x}\theta _{y}+C_{3}},
\end{equation}
where $\mu _{x}$ and $\mu _{y}$ are the average pixel values of image $x$ and $y$ respectively, $\theta _{x}$ and $\theta _{y}$ are the standard deviations of image $x$ and image $y$ respectively, and $\theta _{xy}$ is the covariance of image $x$ and image $y$. $C_{1}$ $\sim$ $C_{3}$ are constants that can maintain stability when denominators are close to 0.

The calculating process of the side-information is shown in Fig. \ref{fig_esi_flowchart}(b). After training the model of the precover estimation network, we can obtain the estimated precover from the input rounded DCT coefficients, then calculate the non-rounded DCT coefficients. The difference between the rounded DCT coefficients and the DCT coefficients without the rounding operation is used as the side-information.
The  details of the CNN based precover estimation architecture are shown in Fig. \ref{fig_esi_network}. It is composed of a series convolution layer (Conv), batch normalization (BN) layer and ReLU layer. All the convolution layers employ 5 $\times$ 5 kernels with stride 1.

\subsubsection{Adjusting the embedding cost}
After obtaining the estimated side-information $\hat{e}$, the embedding cost of Eq. \ref{eq_prob2cost} are further adjusted according to the amplitude and polarity of the estimated side-information $\hat{e}$.

\begin{equation}
\begin{cases}
	\rho _{i,j}^{(ESI)+}=g(\hat{e}_{i,j})\cdot \rho _{i,j}& {\rm if\ } S_{i,j}=C_{i,j}+{\rm sign}(\hat{e}_{i,j})  \\
	\rho _{i,j}^{(ESI)-}=\rho _{i,j}& {\rm if\ } S_{i,j}=C_{i,j}-{\rm sign}(\hat{e}_{i,j})
\end{cases},
\label{cost_adjust}
\end{equation}

\begin{equation}
g(\hat{e}_{i,j})=
\begin{cases}
	1-2\left | \hat{e}_{i,j} \right | & {\rm if\ } \left | \hat{e}_{i,j} \right |\leq \delta \\
	\eta & otherwise
\end{cases},
\label{eq_esi_adjust}
\end{equation}
where $\delta $ and $\eta$ are determined by experiments and will be explained in Section IV. Although \cite{li2019jpeg} tries to use the amplitude of the estimated side-formation, the performance was even inferior because inaccurately estimated side-information will make the performance deteriorate. Here, the amplitude $1-2\left | \hat{e}_{i,j} \right |$ is used for asymmetric adjusting when the absolute value of the side-information is smaller than $\delta$.
We set the parameter $\eta$ to a constant to adjust the cost when the absolute value of the side-information is larger than $\delta$. Because the large estimated side-information may be inaccurate, only the polarity is used for asymmetric embedding to the side-information with a large absolute value.

\section{Experimental Results}

In this section, we will describe the details of the experimental setup and introduce adaptive learning. Then we present the details of adjusting the embedding cost according to the side-information. Finally, we show our experimental results under three different quality factors (QFs) with different payloads.

\subsection{Experimental setup}
The experiments were conducted on SZUBase \cite{tang2017automatic}, BOSSBase v1.01 \cite{bas2011break}, and BOWS2 \cite{bas2007bows}, which contain grayscale images of size $512 \times 512$.
All of the cover images were first resampled to size $256 \times 256$, and the corresponding quantified DCT matrix was obtained using JPEG transformation with quality factor 75.
SZUBase, with 40,000 images, was used to train the JS-GAN and CNN-based precover estimation model.
In the training phase, all parameters of the generator and discriminator were first initialized from a Gaussian distribution.
Then we trained the JS-GAN using the Adam optimizer \cite{kingma2014adam} with a learning rate of 0.0001.
The batch size of the input quantified DCT matrix of the cover image was set to 8.

After a certain number of training iterations, we used the generator and STC encoder \cite{filler2011minimizing} to produce the stego images from 20,000 cover images in BOSSBase and BOWS2 for evaluation.
Four steganalyzers, including the conventional steganalyzers DCTR \cite{holub2014low} and GFR \cite{song2015steganalysis}, as well as the CNN based steganalyzer J-XuNet \cite{xu2017deep}, and SRNet \cite{boroumand2018deep} were used to evaluate the security performance of JS-GAN.
To train the CNN-based steganalyzer, 10,000 cover--stego pairs in BOWS2 and 4,000 pairs in BOSSBase were chosen.
The other 6,000 pairs in BOSSBase were divided into 1,000 pairs as the validation set and 5,000 pairs for testing.
The training stage of JS-GAN was conducted in Tensorflow v1.11 with NVIDIA TITAN Xp GPU card.

\subsection{The content-adaptive learning of JS-GAN}
We trained the JS-GAN for 60 epochs under the target payload of 0.5 bpnzAC and quality factor 75. The best model according to the attack results with GFR was selected. Then stegos with payloads 0.1--0.5 bpnzAC were generated by using STC encoder according to the cost calculated from the best model trained under 0.5 bpnzAC.

\begin{figure}[!htb]
\centering
\subfigure[]{
\begin{minipage}[t]{0.4\linewidth}
\centering
\includegraphics[width=1.3in]{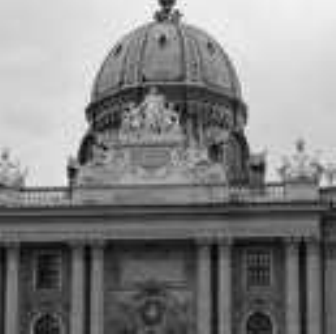}
\end{minipage}
}
\subfigure[]{
\begin{minipage}[t]{0.4\linewidth}
\centering
\includegraphics[width=1.3in]{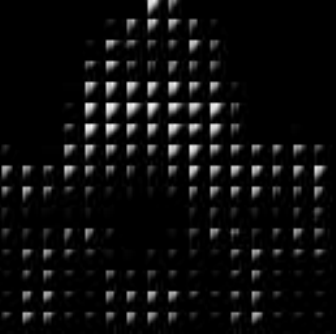}
\end{minipage}
}
\subfigure[]{
\begin{minipage}[t]{0.4\linewidth}
\centering
\includegraphics[width=1.3in]{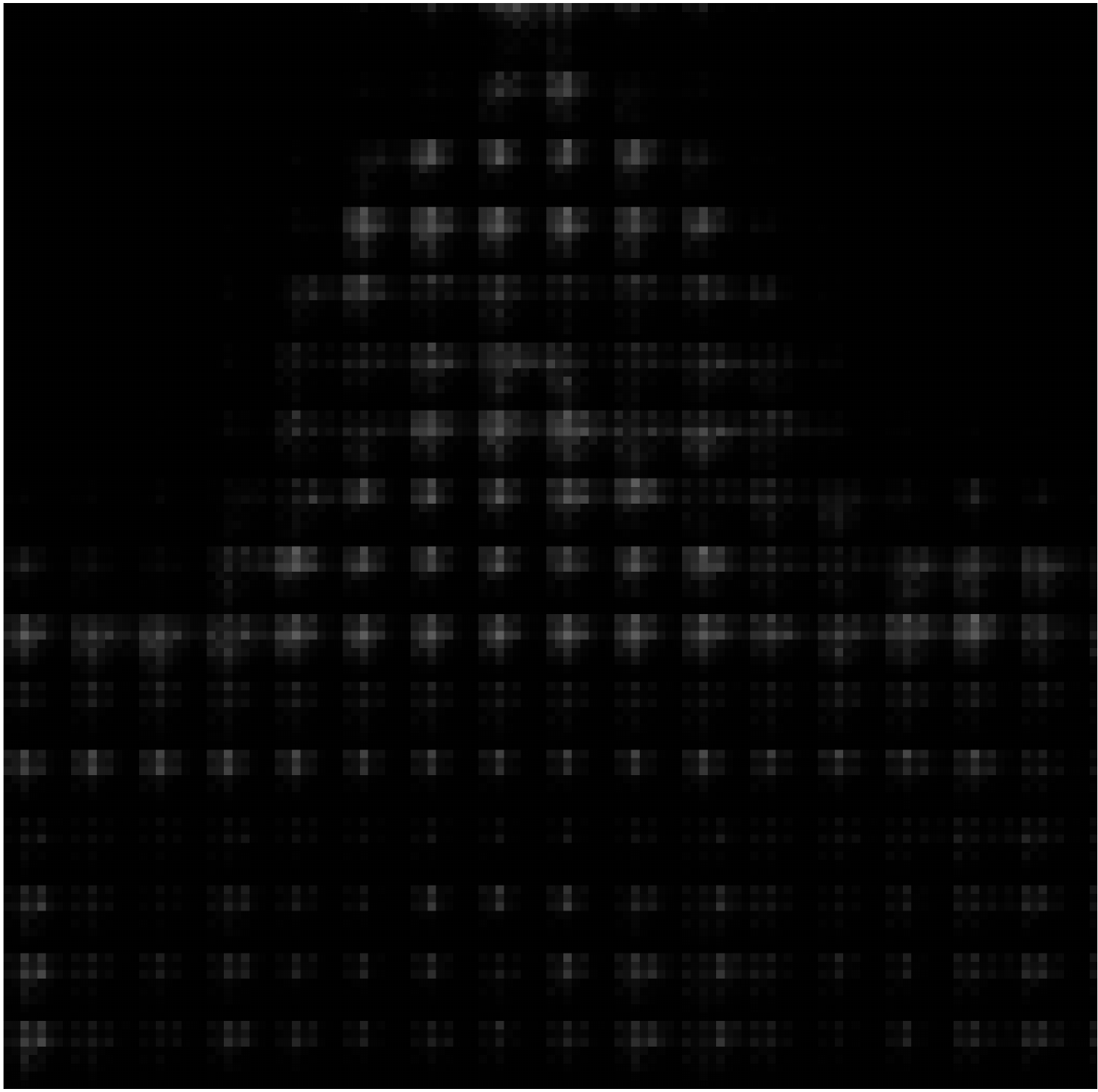}
\end{minipage}
}
\subfigure[]{
\begin{minipage}[t]{0.4\linewidth}
\centering
\includegraphics[width=1.3in]{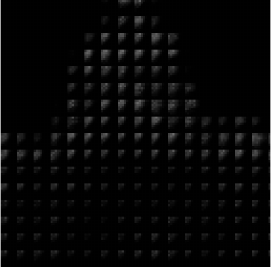}
\end{minipage}
}
\subfigure[]{
\begin{minipage}[t]{0.4\linewidth}
\centering
\includegraphics[width=1.3in]{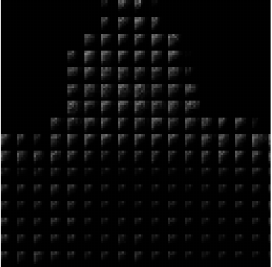}
\end{minipage}
}
\subfigure[]{
\begin{minipage}[t]{0.4\linewidth}
\centering
\includegraphics[width=1.3in]{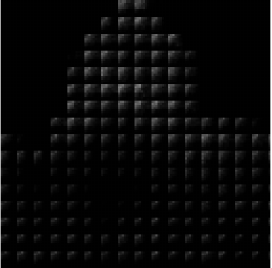}
\end{minipage}
}
\caption{(a) A $128 \times 128 $ crop of BOSSBase cover image ``1013.jpg,'' (b) the embedding probability generated by J-UNIWARD, (c)--(f) the embedding probabilities generated by JS-GAN after 3, 10, 26, and 50 training epochs.}
\label{fig_result_cropped}
\end{figure}

To verify the content-adaptivity of our method, we show the embedding probability map produced by the generator trained after different epochs in Fig.~\ref{fig_result_cropped}.
The embedding probability generated by J-UNIWARD \cite{holub2014universal} with the same payload is also given in Fig.~\ref{fig_result_cropped}(b) for visual comparison.
As shown in Fig.~\ref{fig_result_cropped}, the $8 \times 8$ block property of the embedding probability has been automatically learned.
The global adaptability (inter-block) and the local adaptability (intra-block) have improved with an increase in  the number of training epochs.
After 50 training epochs, the block in the complex region has a large embedding probability, which verifies the global adaptability.
Inside the $8 \times 8$ block, the larger probabilities are mostly located in the top left low frequency region, which proves the local adaptability of JS-GAN.
It can be also seen that the embedding probability generated by our GAN-based method has similar characteristics to those generated by J-UNIWARD.

\subsection{Parameter selection for adjusting the embedding cost }
\begin {table*}[!htbp]
\renewcommand\arraystretch{1.2}
\caption {Error rates of JS-GAN(ESI) with 0.4 bpnzAC detected by GFR under different values of $\eta$ ($\delta=0$)}
\label{tab:select eta}
\begin{center}
\begin{tabular}{ |c|c|c|c|c|c|c|c| }
	\hline
	$\eta$    	&0.55	   & 0.6	 & 0.65	    & 0.7	    & 0.75          & 0.8    & 1   \\
	\hline
	error rate  &0.2554	  & 0.2607   &\textbf{0.2785}  	&0.2784    	&0.2758         &0.2622  & 0.2 \\ 		 			
	\hline		
\end{tabular}
\end{center}
\end{table*}

\begin {table*}[!htbp]
\renewcommand\arraystretch{1.2}
\caption {Error rates of JS-GAN(ESI) with 0.4 bpnzAC detected by GFR under different values of $\delta$ ($\eta=0.65$)}
\label{tab:select delta}
\begin{center}
\begin{tabular}{ |c|c|c|c|c|c|c|c|c| }
\hline

$\delta$    &0	           &0.01	 &0.05	         &0.1	    &0.15	    &0.2	     &0.4	    &0.5	 \\
\hline
error rate  & 0.2785	   &0.2829  &\textbf{0.3125} & 0.2969	& 0.2944	& 	0.2954   & 0.2799	& 0.2132 \\
\hline

\end{tabular}
\end{center}
\end{table*}

\begin{table*}[htb]
	\centering
\caption{Error rates detected by different steganalyzers when QF = 75}
\begin{tabular}{|c|c|c|c|c|c|c|}
\hline
\multirow{2}{*}{Steganalyzer} & \multirow{2}{*}{Steganographic scheme} & \multicolumn{5}{c|}{Payload}                                   \\ \cline{3-7}
&                                        & 0.1 bpnzAC & 0.2 bpnzAC & 0.3 bpnzAC & 0.4 bpnzAC & 0.5 bpnzAC \\ \hline
\multirow{4}{*}{GFR}          & JS-GAN(ESI)                            & \textbf{0.4720}     & \textbf{0.4220}     &\textbf{0.3707}     & \textbf{0.3125}     & \textbf{0.2537}     \\ \cline{2-7}
& JS-GAN                                & 0.4400     & 0.3650     & 0.2826     & 0.2000     & 0.1283     \\ \cline{2-7}
& UERD                                   & 0.4273     & 0.3351     & 0.2468     & 0.1749     & 0.1126     \\ \cline{2-7}
& J-UNIWARD                              & 0.4393     & 0.3488     & 0.2568     & 0.1742     & 0.1090     \\ \hline

\multirow{4}{*}{DCTR}         & JS-GAN(ESI)                            & \textbf{0.4799}     & \textbf{0.4474}     & \textbf{0.4084}     & \textbf{0.3529}    & \textbf{0.2848}         \\ \cline{2-7}
& JS-GAN                                & 0.4572     & 0.3916     & 0.3123     & 0.2303     & 0.1528     \\ \cline{2-7}
& UERD                                   & 0.4575     & 0.3871     & 0.3137     & 0.2390     & 0.1690     \\ \cline{2-7}
& J-UNIWARD                              & 0.4655     & 0.4011     & 0.3238     & 0.2492     & 0.1701     \\ \hline

\multirow{4}{*}{J-XuNet}      & JS-GAN(ESI)                            & \textbf{0.4053}     & \textbf{0.3030}      & \textbf{0.2232}     & \textbf{0.1622}     & \textbf{0.1147}          \\ \cline{2-7}
& JS-GAN                                & 0.4079     & 0.2873     & 0.1859     & 0.1331     & 0.0834          \\ \cline{2-7}
& UERD                                   & 0.3256     & 0.1923     & 0.1122     & 0.0709     & 0.0459     \\ \cline{2-7}
& J-UNIWARD                              & 0.3977     & 0.2750     & 0.1821     & 0.1166     & 0.0752     \\ \hline

\multirow{4}{*}{SRNet}       & JS-GAN(ESI)                            &\textbf{0.3268} 	 &\textbf{0.2142} 	&\textbf{0.1417} 	&\textbf{0.0902} 	&\textbf{0.0618}          \\ \cline{2-7}
& JS-GAN                                &0.3085 	 &0.1867 	&0.1114 	&0.0640 	&0.0292          \\ \cline{2-7}
& UERD                                   & 0.1949  & 0.0865   & 0.0446   & 0.0311     & 0.0143     \\ \cline{2-7}
& J-UNIWARD                              & 0.2971  & 0.1771   & 0.1016   & 0.0596     & 0.0311     \\ \hline
\end{tabular}
\label{table_QF_75}
\end{table*}

We conducted experiments with JS-GAN(ESI) with 0.4 bpnzAC and quality factor of 75 to select the proper parameters for asymmetric embedding.
The error rate $P_{E}$ of the steganalyzers is used to quantify the security performance of our proposed framework.
The error rate detected by GFR with an ensemble classifier is used to evaluate the performance.
Firstly, we investigate the impact of the sign of the estimated side-information $\hat{e}$.
Thus we set the amplitude parameter $\delta = 0$ and observe the effect of the polarity parameter $\eta$ in Eq. \ref{eq_esi_adjust}.

The effect of $\eta$ is shown in Table \ref{tab:select eta}, $\eta= 1$ denotes the original JS-GAN, which the cost of +1 is equal to the cost -1.
Experimental results show that the performance would be improved when $\eta$ decreases. Thus the performance can be improved by adjusting the cost asymmetrically according to the sign of the estimated side-information.
From Table \ref{tab:select eta}, the detection error rate of JS-GAN(ESI) increases 7.85\% with $\eta=0.65$ compared with the original JS-GAN.

After putting $\eta=0.65$, we further investigated the influence of the parameter $\delta$ by using the polarity and the amplitude of the side-information at the same time.
It can be seen from Table \ref{tab:select delta} that setting $\delta=0.05$ will achieve better performance than other values, it also can improve the performance about 10\% over setting $\delta=0.5$, which was used in \cite{li2019jpeg}.
The experimental results show that only a small amplitude can be used to improve the security performance. This is because amplitudes close to 0.5 are not precise. It can also be seen that setting $\delta=0.05$ will lead to an improvement by about 3.4\% over $\delta=0$, and this shows that using the amplitude $1-2\left | \hat{e}_{i,j} \right |$ properly can further improve the performance over using only the polarity of the side-information.
From Table \ref{tab:select eta} and Table \ref{tab:select delta}, we set $\eta=0.65$ and $\delta=0.05$ for the final version JS-GAN(ESI).

\begin{figure*}[!t]
	\centering
	\subfigure[]{
		\begin{minipage}[t]{0.4\linewidth}
			\centering
			\includegraphics[width=2.6in]{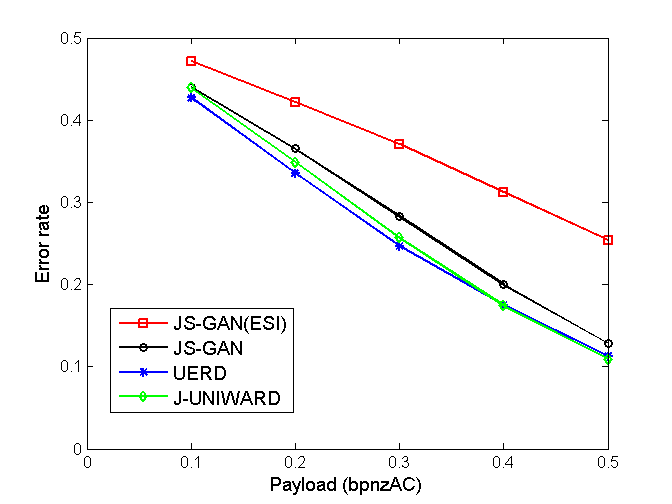}
		\end{minipage}
	}
	\subfigure[]{
		\begin{minipage}[t]{0.4\linewidth}
			\centering
			\includegraphics[width=2.6in]{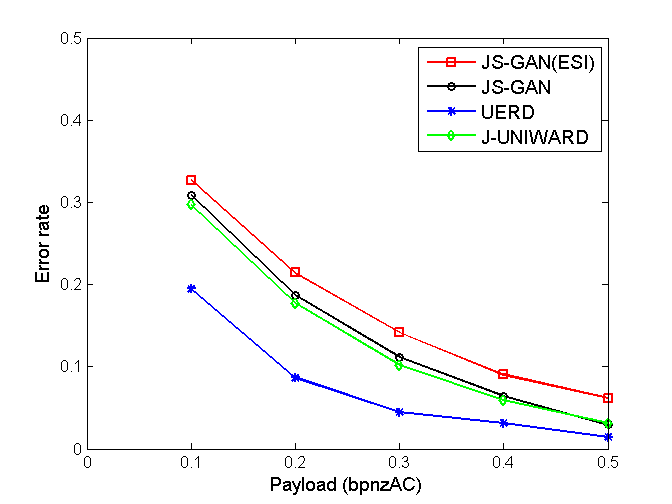}
		\end{minipage}
	}

	\caption{Error rates detected by different steganalyzers when QF = 75. (a) Detected by GFR, (b) Detected by SRNet.  }
	\label{QF75err}
\end{figure*}

\begin{table*}[!htb]
	\centering
\caption{Error rates detected by different steganalyzers when QF = 50}
\begin{tabular}{|c|c|c|c|c|c|c|}
\hline
\multirow{2}{*}{Steganalyzer} & \multirow{2}{*}{Steganographic scheme} & \multicolumn{5}{c|}{Payload}                                   \\ \cline{3-7}
&                                        & 0.1 bpnzAC & 0.2 bpnzAC & 0.3 bpnzAC & 0.4 bpnzAC & 0.5 bpnzAC \\ \hline
\multirow{4}{*}{GFR}          & JS-GAN(ESI)                           &\textbf{0.4596} 	&\textbf{0.4067} 	&\textbf{0.3392} 	&\textbf{0.2730} 	&\textbf{0.2119}
\\ \cline{2-7}
& JS-GAN                               &0.4222 	&0.3243 	&0.2322 	&0.1524 	&0.0937
\\ \cline{2-7}
& UERD                                   & 0.4110     & 0.3100     & 0.2181    & 0.1421    & 0.0899     \\ \cline{2-7}
& J-UNIWARD                              &0.4151     & 0.3104     & 0.2029     &0.1277     &0.0735     \\ \hline

\multirow{4}{*}{DCTR}         & JS-GAN(ESI)                            &\textbf{0.4688}     & \textbf{0.4210}     & \textbf{0.3604}     &\textbf{0.3036}    & \textbf{0.2483}         \\ \cline{2-7}
& JS-GAN                                & 0.4304    & 0.3360     & 0.2431     & 0.1620   &0.0982     \\ \cline{2-7}
& UERD                                   &0.4363     & 0.3560     & 0.2696     &0.1918    & 0.1287      \\ \cline{2-7}
& J-UNIWARD                              &0.4386     & 0.3571     & 0.2710     &0.1837    & 0.1211      \\ \hline

\multirow{4}{*}{J-XuNet}      & JS-GAN(ESI)                           &\textbf{0.4049}     & \textbf{0.2990}     & \textbf{0.2144}     &\textbf{0.1563}    & \textbf{0.1104}          \\ \cline{2-7}
& JS-GAN                               &0.3701     & 0.2466     & 0.1589     &0.0977    & 0.0647          \\ \cline{2-7}
& UERD                                  &0.2983     & 0.1630     & 0.0952     &0.0554    & 0.0345     \\ \cline{2-7}
& J-UNIWARD                             &0.3764     & 0.2361     & 0.1441     &0.0861    & 0.0514      \\ \hline

\multirow{4}{*}{SRNet}       & JS-GAN(ESI)                         &\textbf{0.3266}     & \textbf{0.2154}     & \textbf{0.1262}     &\textbf{0.0804}    & \textbf{0.0474}    \\ \cline{2-7}
& JS-GAN                             &0.2702     & 0.1491     & 0.0781     &0.0377    & 0.0188    \\ \cline{2-7}
& UERD                                &0.1764     & 0.0782     & 0.0408     &0.0182    & 0.0178    \\ \cline{2-7}
& J-UNIWARD                           &0.2808     & 0.1402     & 0.0732     &0.0377    & 0.0334    \\ \hline
\end{tabular}
\label{table_QF_50}
\end{table*}

\begin{table*}[!htb]
	\centering
\caption{Error rates detected by different steganalyzers when QF = 95}
\begin{tabular}{|c|c|c|c|c|c|c|}
\hline
\multirow{2}{*}{Steganalyzer} & \multirow{2}{*}{Steganographic scheme} & \multicolumn{5}{c|}{Payload}                                   \\ \cline{3-7}
&                                        & 0.1 bpnzAC & 0.2 bpnzAC & 0.3 bpnzAC & 0.4 bpnzAC & 0.5 bpnzAC \\ \hline
\multirow{4}{*}{GFR}          & JS-GAN(ESI)                            & 0.4817     & 0.4525     & \textbf{0.4122}     & \textbf{0.3652}     & \textbf{0.3095}     \\ \cline{2-7}
& JS-GAN                                & 0.4783     & 0.4434     & 0.3920     & 0.3382     & 0.2634     \\ \cline{2-7}
& UERD                                   & 0.4754     & 0.4268     & 0.3702     & 0.3072     & 0.2423     \\ \cline{2-7}
& J-UNIWARD                              & \textbf{0.4894}     & \textbf{0.4549}     & 0.4033     & 0.3417     & 0.2791     \\ \hline
\multirow{4}{*}{DCTR}         & JS-GAN(ESI)                            & 0.4868     & 0.4653     & 0.4295     & 0.3859     & 0.3348          \\ \cline{2-7}
& JS-GAN                                & 0.4864     & 0.4599     & 0.4150     & 0.3604     & 0.2877          \\ \cline{2-7}
& UERD                                   & 0.4888     & 0.4597     & 0.4261     & 0.3697     & 0.3093     \\ \cline{2-7}
& J-UNIWARD                              & \textbf{0.4953}     & \textbf{0.4750}     & \textbf{0.4450}     & \textbf{0.3984}     & \textbf{0.3430}     \\ \hline

\multirow{4}{*}{J-XuNet}      & JS-GAN(ESI)                            & 0.4706     & 0.4118     & 0.3340     & 0.2615     & 0.1811          \\ \cline{2-7}
& JS-GAN                                & 0.4752     & 0.4258     & 0.3441     & 0.2735     & 0.1893          \\ \cline{2-7}
& UERD                                   & 0.4295     & 0.3358     & 0.2380     & 0.1773     & 0.1174     \\ \cline{2-7}
& J-UNIWARD                              & \textbf{0.4841}     & \textbf{0.4434}     & \textbf{0.3988}     & \textbf{0.3309}     & \textbf{0.2689}     \\ \hline

\multirow{4}{*}{SRNet}       & JS-GAN(ESI)  & 0.4344 & 0.3125 &0.2103 &0.1364 &0.0779          \\ \cline{2-7}
& JS-GAN  &  0.4267 &0.3043 &0.2062 &0.1286 &0.0723
\\ \cline{2-7}
& UERD   & 0.3201     & 0.1980     & 0.1167     & 0.0673     & 0.0478     \\ \cline{2-7}
& J-UNIWARD   & \textbf{0.4349}     & \textbf{0.3250}     & \textbf{0.2347}     & \textbf{0.1599}     & \textbf{0.1008}     \\ \hline
\end{tabular}
\label{table_QF_95}
\end{table*}

\subsection{Results and analysis}

We selected the classic steganographic algorithms UERD and J-UNIWARD to make a comparison with our proposed methods JS-GAN and JS-GAN(ESI). For a fair comparison, all steganographic algorithms used the STC encoder to embed the messages. Four different steganalyzers were used to evaluate the performance, include the conventional steganalyzers GFR and DCTR, as well as the CNN-based steganalyzers J-XuNet and SRNet.

The experimental results for JPEG compressed images with quality factor 75 are shown in Table \ref{table_QF_75} and Fig. \ref{QF75err}.
 From Table \ref{table_QF_75}, it can be seen that under the attack of the conventional steganalyzer GFR, our proposed architecture JS-GAN can obtain better security performance than the conventional methods UERD and J-UNIWARD. For example, under the attack of GFR with 0.4 bpnzAC, the JS-GAN can achieve a performance better by 2.51\% and 2.58\% than UERD and J-UNIWARD respectively. Under the attack of the CNN based steganalyzer SRNet with 0.4 bpnzAC, the proposed JS-GAN can increase the detection error rate by 3.29\% and 0.44\% over UERD and J-UNIWARD respectively.

It can be seen from Table \ref{table_QF_75} that the security performance of JS-GAN can be further improved significantly by incorporating the proposed side information estimated method. With the payload 0.4 bpnzAC, the JS-GAN(ESI) can increase the detection error rate by 11.25\% over JS-GAN against the conventional steganlayzer GFR, and it can also increase the detection error rate by 2.62\% over JS-GAN to against the attack of the CNN based steganlayzer SRNet. The main reason for the high level of security achieved is mainly making the embedding preserve the statistical characteristics of the precover.

Compared with the conventional methods UERD and J-UNIWARD, a significant improvement can be achieved by combining the proposed JS-GAN with estimated side-information. Under the detection of GFR with the payload 0.4 bpnzAC, the detection error of JS-GAN(ESI) is increased by 13.76\% and 13.83\% over that of UERD and J-UNIWARD respectively. Under the detection of SRNet with the payload 0.4 bpnzAC, the detection error of JS-GAN(ESI) is also up by 5.91\% and 3.06\% over UERD and J-UNIWARD respectively.

To verify the proposed method in the face of different compression quality factors, we conducted the experiments for JPEG compressed images with quality factors 50 and 95. The parameters of the generator and discriminator with the quality factors 50 and 95 were initialized from the trained best model using the quality factor 75.
The experimental results are shown in Table \ref{table_QF_50} and Table \ref{table_QF_95} respectively.
From Table \ref{table_QF_50}, JS-GAN(ESI) achieves significant improvement compared with the other three steganographic algorithms under the attack of different steganalyzers with different payloads, which proves that the proposed method can achieve better security performance.
From Table \ref{table_QF_95}, it can be seen that JS-GAN(ESI) can achieve comparable performance with other steganographic algorithms, and it can also be seen that the performance of JS-GAN(ESI) for higher quality factor needs to be further improved.

\section{Conclusion}

In this paper, an automatically embedding cost function learning framework called JS-GAN has been proposed for JPEG steganography. Our experimental results obtained by conventional and CNN based steganalyzers allow us to draw the following conclusions:
\begin{enumerate}[1)]
 \item Through an adversarial training between the generator and discriminator, an automatic content-adaptive steganography algorithm can be designed for the JPEG domain. Compared with the conventional heuristically designed JPEG steganography algorithms, the automatic learning method can adjust the embedding strategy flexibly according to the characteristics of the steganalyzer.
 \item To simulate the message embedding and avoid the gradient-vanishing problem, a gradient-descent friendly and highly efficient probability based embedding simulator can be designed. Experimental results show that the probability-based embedding simulator can make a contribution to learn the local and global adaptivity.

  \item When the original uncompressed image is not available, a properly trained CNN based side-information estimation model can acquire the estimated side-information. The security performance can be further improved dramatically by using asymmetric embedding with the well estimated side-information.
\end{enumerate}

This paper opens up a promising direction in embedding cost learning in the JPEG domain by adversarial training between the generator and discriminator. It also shows that predicting the side-information brings a significant improvement when only the JPEG compressed image is available. Further investigation could include the following aspects. Firstly, higher efficiency generators and discriminators can be further investigated. Secondly, the conflict with the gradient-vanishing and simulating the embedding process also have much room for improvement. Thirdly, we have only carried out an initial study using the CNN to estimate the side-information, and the asymmetric adjusting strategy deserves further research. Finally, we will further improve the performance under the larger quality factor 95.

\bibliographystyle{IEEEtran}

\bibliography{JS_GAN}

\end{document}